\documentclass[aps,prd, superscriptaddress,showpacs,nofootinbib]{revtex4}
\usepackage{amsmath,amssymb}

\usepackage{graphicx}

\newcommand{\be}{\begin{eqnarray}}
\newcommand{\ee}{\end{eqnarray}}
\newcommand{\ba}{\begin{eqnarray}}
\newcommand{\ea}{\end{eqnarray}}

\def\ben{\begin{equation}}
\def\een{\end{equation}}
\def\bena{\begin{eqnarray}}
\def\eena{\end{eqnarray}}

\begin{document}

\title{On the Covariant Galileon and a consistent self-accelerating Universe}

\author{Cristiano Germani}
\email{cristiano.germani@lmu.de}
\affiliation{Arnold Sommerfeld Center, Ludwig-Maximilians-University, Theresienstr. 37, 80333 Muenchen, Germany}

\begin{abstract}
In this paper we show that the flat space Galilean theories with up to three scalars in the equation of motion (the quartic Galileons) are recovered in the decoupling limit of certain scalar theories non-minimally coupled to gravity, the so-called ``Slotheonic" theories. These theories are also invariant under the generalized Galilean shifts in curved spacetime. While Galilean self-(derivative)couplings are not explicit in the action, they appear after integrating out gravity.  
We then argue that Galilean supersymmetric theories may only be found in the context of supergravity. Finally, we discuss on the possibility that Slotheonic theories are the effective four dimensional theories of consistent DGP-like models with self-accelerating cosmological solutions. Moreover, we show that the quartic and cubic Galileon in consistent DGP models cannot be decoupled.
\end{abstract}


\maketitle
 \section{Galilean theories}

A massless scalar field ($\pi$), in flat spacetime and in cartesian coordinates, is described by a purely second order equation, the Klein-Gordon equation
\be\label{kg}
\partial^2_t\pi-\partial^2_x\pi=0\ .
\ee
From (\ref{kg}) it is easy to see that the scalar is invariant under the Galilean symmetry
\be\label{gal}
\pi\rightarrow \pi+a+b_\mu x^\mu\ ,
\ee
where $a$ and $b_\mu$ are respectively a constant and a constant vector. 

The Galilean symmetry (\ref{gal}) may be extended on non-cartesian coordinates by considering covariantly constant Killing vectors (labeled by $a,b\ldots$) $\xi_\alpha^a$ as \cite{sloth}
\be\label{galcov}
\pi\rightarrow \pi+a+b_a\int \xi_\alpha^a dx^\alpha\ ,
\ee
where, as we said, $\nabla_\mu\xi_\nu^a=0$. The formulation (\ref{galcov}) allows us to extend the Galilean symmetry to curved spacetime \cite{sloth}.

Any theory of a scalar field with equation of motion involving only second or higher order covariant derivatives, is obviously bound to be invariant under the symmetry (\ref{galcov}). What is non trivial, is to find Lagrangian theories that involve {\it only} second order derivatives at the equation of motion level. This last requirement is important to avoid possible Ostrogardski instabilities \cite{ostro}. 

Motivated by the decoupling limit of the Dvali-Gabadadze-Porrati (DGP) model \cite{dgp} it has been noticed \cite{galileo} that indeed a set of self-coupled, second order in derivatives, scalar field theories encoding the same Galilean symmetry (\ref{gal}) of a free massless scalar field, do exist. Such theories are dubbed Galilean theories. 
Galilean theories are subclass of the general tensor-scalar theories involving only up to second order derivatives originally found by Hordenski \cite{hor}.

The Galilean theories are classified in terms of the number of field $(n+1)$ appearing at the Lagrangian level. These theories produce the following equations of motion \cite{galileo}
\be\label{flateqm}
\sum_{n=1}^4c_{n}{\cal E}_{n+1}=0\ ,
\ee
where
\begin{eqnarray}
{\cal E}_{1}&=&1\cr
{\cal E}_2&=&\square\pi\cr
{\cal E}_3&=&(\square\pi)^2-(\partial_{\mu\nu}\pi)^2\cr
{\cal E}_4&=&(\square\pi)^3-3\square\pi(\partial_{\mu\nu}\pi)^2+2(\partial_{\mu\nu}\pi)^3\cr
{\cal E}_5&=&(\square\pi)^4-6(\square\pi)^2(\partial_{\mu\nu}\pi)^2+8\square\pi(\partial_{\mu\nu}\pi)^3+3\left[(\partial_{\mu\nu}\pi)^2\right]^2-6(\partial_{\mu\nu}\pi)^4\ ,
\end{eqnarray}
$(\partial_{\alpha\beta}\pi)^n$ denotes the cyclic contractions and finally, $c_n$ are constants.

The same equations may be also written in a simpler form \cite{simpler} for $0<n\leq 4$
\begin{eqnarray}
{\cal E}_{n+1}=\frac{1}{(4-n)!}\epsilon^{\mu_1\mu_3\ldots\mu_{2n-1}\ \nu_1\ldots\nu_{4-n}}\epsilon^{\mu_2\mu_4\ldots\mu_{2n}}{}_{\nu_1\ldots\nu_{4-n}}\pi_{\mu_1 \mu_2}\ldots\pi_{\mu_{2n-1}\mu_{2n}}\ ,
\end{eqnarray}
where $\pi_\mu\equiv\nabla_\mu\pi$ and the volume form
\be
\epsilon^{\mu_1 \mu_2\mu_3\mu_4}=-\frac{\delta_1^{[\mu_1}\delta_2^{\mu_2}\delta_3^{\mu_3}\delta_4^{\mu_4]}}{\sqrt{-g}}\ .
\ee

 \subsection{Curved spacetime}
 
Until this point we have only discussed Galilean symmetric scalar field theories in flat spacetime. However, we already noticed that the Galilean symmetry is really a shift along a covariantly constant Killing directions. It is therefore easy to extend this symmetry to curved spacetimes containing a set of covariantly constant Killing forms $\xi_\mu^a$ \cite{sloth}, or, more technically, spacetimes with closed forms $d\xi^a=0$. 

Spacetimes admitting integrable Killing vectors are of particular type \cite{Stefani}.  A  Killing vector $\xi^\mu$ can be covariantly constant only if $\xi$ satisfies the algebraic condition  
\be\label{ruin}
R^{\mu}{}_{\nu\rho\sigma}\xi^\nu=0\ ,
\ee
which can be obtained from the consistency condition $[\nabla_\rho,\nabla_\sigma]\xi^\mu=0$.  In other words, the holonomy group of space-time must be reduced to a subgroup of SO$(1,3)$. Explicitly, if the 
vector is non-null, the space-time metric is of the form
\be\label{st1}
d s^2=g_{ij}(x^k) d x^i d x^j +\kappa\, d y^2\, , ~~~i,j,k=1,2,3\ ,
\ee
where $\kappa=+1,-1$ for spacelike or timelike $\xi^\mu$, respectively, or for a null $\xi^\mu$
\be\label{st2}
d s^2=g_{ij}(x^k) d x^i d x^j +d z d y\, , ~~~i,j,k=1,2,3\ ,
\ee
where $z$ is any coordinate in the $i$'s directions.

It has been proven in \cite{sloth} that all theories invariant under (\ref{galcov}), in non-trivial spacetimes with integrable Killing vectors, and producing only second order derivatives in the equations of motion, can be obtained from a linear combinations of the following Lagrangians
\begin{eqnarray}\label{curved}
{\cal L }_0&=&\frac{M_p^2}{2}R\ ,\cr
{\cal L }_{1}&=& M_0^3 \pi+M_1 \pi R+\frac{\pi}{M_{2}} GB\ ,\cr
{\cal L }_{2}&=&-\frac{1}{2}g^{\mu\nu}\partial_\mu\pi\partial_\nu\pi+\frac{1}{2 M_3^2}G^{\mu\nu}\partial_\mu\pi\partial_\nu\pi\ ,\cr
{\cal L }_{3}&=&\frac{1}{M_4^3}(\partial\pi)^2\square\pi\ ,
\end{eqnarray}
and finally,
\be\label{ex}
{\cal L}_{\rm extra}=\frac{1}{M_5^5} {}^{**}R^{\alpha\beta\mu\nu}\partial_\alpha\pi\partial_\mu\pi\nabla_\beta\nabla_\nu\pi\ .
\ee
where $GB=R_{\alpha\beta\gamma\delta}R^{\alpha\beta\gamma\delta}-4R_{\alpha\beta}R^{\alpha\beta}+R^2$ is the Gauss-Bonnet combination, $M_i$ are some mass scales and $M_p$ is the Planck scale. The sign of the terms in ${\cal L}_2$ are chosen in such a way that, whenever energy conditions are satisfied, the effective propagator of $\pi$ is never ghost-like \cite{sloth}. The double dual Riemann tensor is defined as 
\be
{}^{**}R^{\mu_1\mu_2\nu_1\nu_2}\equiv-\frac{1}{4}\epsilon^{\mu_1 \mu_2\mu_3\mu_4}~
\epsilon^{\nu_1 \nu_2\nu_3\nu_4}R_{\mu_3\mu_4\nu_3 \nu_4}\ .
\ee
We have isolated the the double dual Riemann tensor Lagrangian (\ref{ex}) as, as we shall show, the dimensional reduction of a consistent (five dimensional) DGP model, does not contain this term. Note that instead in the dimensional reduction of the original DGP model \cite{dgp}, the Lagrangian (\ref{ex}) would also be present \cite{claudia}.

Interestingly enough, the fact that (\ref{ex}) is invariant under the curved spacetime Galilean transformation, automatically implies that, in the spacetimes (\ref{st1},\ref{st2}) $GB\equiv 0$, or, in turn, that the Euler characteristic of these spacetimes must vanish. The proof comes from noticing that, with the help of the Gauss-Bonnet identity in four dimensions
\be
\frac{\delta GB}{\delta g^{\alpha\beta}}\equiv 0\ ,
\ee 
the equation of motion coming from the scalar variation of (\ref{ex}) reads \cite{fabfour}
\be
\frac{\delta{\cal L}_{\rm extra}}{\delta\pi}=\ldots+\pi_\alpha\pi^\alpha GB\ .
\ee 
We then directly see that, since this action is invariant under Galilean symmetry, the Gauss-Bonnet combination must vanish. This can be also proven by direct computation from the metric (\ref{st1},\ref{st2}).

While ${\cal E}_{1,2,3}$ are straightforwardly obtained as a $\pi$ variation of ${\cal L}_{1,2,3}$ in the decoupling limit of vanishing curvatures, the same is not true for ${\cal E}_{4,5}$. We will then show that these two equations are in fact special. In particular, ${\cal E}_4$ will be a {\it derived} equation from some combination of (\ref{curved}) in the gravity decoupling limit. 

Let us finally mention that the covariantization of the Lagrangians leading to ${\cal E}_{1,2,3}$ is straightforward. Indeed, one may just minimally covariantize these theories by substituting, in the Lagrangians generating ${\cal E}_{1,2,3}$, any partial derivative into a covariant one. However, as it has been noticed in \cite{covgal}, the straightforward covariantization of ${\cal E}_{4,5}$, would produce higher derivatives in the equations of motion. This is mainly due to the non-commutativity of derivatives acting to the scalars in curved spacetime. In order to solve this problem, counter terms involving non-minimal couplings of the scalar derivatives to curvatures have been added in the so-called covariant Galileon \cite{covgal}. We will show here that these counter terms that covariantize consistently ${\cal E}_4$ are hidden in ${\cal L}_{0,2}$, once gravity, via the Einstein equations, is integrated out in the scalar field equation. 

The main difference from the quartic to the quintic Galileon is that the scalar equation that can potentially generate ${\cal E}_5$ via integrating out gravity, involves the full Riemann tensor via the action (\ref{ex}). In this case then, Einstein equations cannot be straightforwardly used to integrate out gravity. It is however interesting to note that a consistent dimensional reduction of the DGP model would not contain (\ref{ex}). This is leading us to conjecture that the quintic Galileon may indeed not be obtained throughout the theory (\ref{curved},\ref{ex}) and therefore in turn, that the flat Galilean invariance of the quintic Galileon cannot be extended to curved spacetime. Or, in other words, any consistent flat limit of any combinations of the actions (\ref{curved},\ref{ex}) should always lead to a vanishing contribution of (\ref{ex}). The proof of this is however postponed for future work.

\section{Covariant quartic Galileon is Slotheon}
 
Among all theories invariant under the extended Galilean symmetry (\ref{galcov}) in curved spacetime, the combination of ${\cal L}_0$ and ${\cal L}_2$ as been dubbed in \cite{sloth} as the ``Slotheonic" theory. The name reflect the fact that the Slotheon, is a scalar field with lower magnitude of kinetic energy with respect to the same theory with only minimal couplings. The Slotheon is indeed the base of what is called ``high-friction" inflation \cite{high} which is the mechanism under which both the Standard Model Higgs boson \cite{new} and an axion \cite{uv}, can produce a reliable, successful, inflation.
Finally, the Slotheonic theory can also be supersymmetrized in the context of new-minimal supergravity \cite{sugra}.

The variation of the Slotheonic theory under $\pi$ gives the following equation
\be\label{door}
\pi^\alpha{}_\alpha-\frac{1}{M_3^2}G^{\alpha\beta}\pi_{\alpha\beta}=0\ ,
\ee 
while the variation with respect to the metric leads to the following Einstein equations \cite{klu}
\be\label{G}
G_{\mu\nu}=M_{\rm P}^{-2} T_{\mu\nu}\ ,
\ea
where
\be\label{T}
T_{\mu\nu}=\pi_\mu\pi_\nu-\frac{1}{2}g_{\mu\nu}(\partial\pi)^2+\frac{\Theta_{\mu\nu}}{M_3^2}\ ,
\ee
and
\be
\Theta_{\mu\nu}=\frac{1}{2}\pi_{\mu}\pi_{\nu}R-2\pi_{\alpha}\pi_{(\mu}R^{\alpha}_{\nu)}+\frac{1}{2}\pi_{\alpha}\pi^{\alpha}G_{\mu\nu}-\pi^{\alpha}\pi^{\beta}R_{\mu\alpha\nu\beta}-\pi_{\alpha\mu}\pi^{\alpha}_\nu+\pi_{\mu\nu}\pi_\alpha^{~\alpha}+\frac{1}{2}g_{\mu\nu}[\pi_{\alpha\beta}\pi^{\alpha\beta}-(\pi_\alpha^{~\alpha})^2+2\pi_\alpha\pi_\beta R^{\alpha\beta}]\ .\nonumber
\ee
It is already clear that the zero curvature parts of $\Theta_{\mu\nu}$ contain the seeds of the flat spacetime quartic Galileon.

A consistent flat space limit is obtained by taking $M_p\rightarrow \infty$ such that $G_{\alpha\beta}\rightarrow 0$. This would simply single out the standard Klein-Gordon term $\square \pi=0$. 

What we are interested in, however, is a set of parameter such that $G_{\alpha\beta}\rightarrow 0$, in order to obtain a consistent flat limit, but at the same time $\frac{1}{M_3^2}G_{\alpha\beta}$ finite, in order to obtain a non-trivial scalar field equation. This is what we are going to discuss in the following. Since we want to focus on the higher order Galilean theories we will not consider the canonical term of the scalar field any longer in this section.

\subsection{Recovering ${\cal E}_{4}$}

In flat spacetime, the Galilean equations of motion (\ref{flateqm}) involve up to four scalars. In curved spacetime however, the variation of (\ref{curved}) with respect to $\pi$ would instead explicitly only involve a maximum of $1$ scalar. Therefore, if higher order Galilean theories should appear, they can only be obtained by integrating out gravity. 

It is indeed easy to see that the limit $M_p\rightarrow \infty$ together with $\Lambda_4\equiv (M_p M_3^2)^{1/3}\rightarrow {\rm finite}$ would reproduce ${\cal E}_4$ from the only use of ${\cal L}_{0,2}$\footnote{Whenever a Lagrangian of (\ref{curved}) is not explicitly mentioned it is considered vanishing via suitable limit of parameters.}. 

First of all these limits generate the following hierarchy of scales
\be
M_p\gg \sqrt{M_3 M_p}\gg \left(M_3^2 M_p\right)^{1/3}\ .
\ee
From (\ref{G}) we then see, by an iterative substitution of curvatures from the Einstein equations, that
\be
G_{\mu\nu}=\frac{-\pi_{\alpha\mu}\pi^{\alpha}_\nu+\pi_{\mu\nu}\pi_\alpha^{~\alpha}+\frac{1}{2}g_{\mu\nu}[\pi_{\alpha\beta}\pi^{\alpha\beta}-(\pi_\alpha^{~\alpha})^2]}{M_3^2 M_p^2}+{\cal O}{\left(\frac{1}{M_p^4 M_3^4}\right)}\ .
\ee
The above expansion then leads to the following desired result\footnote{Note that although $M_p M_3^2\rightarrow {\rm finite}$, $M_p^4 M_3^6=\Lambda_4^9 M_p\rightarrow \infty$.}
\begin{eqnarray}\label{similar}
\lim_{\substack{\tiny M_p\rightarrow \infty\\ \Lambda_4\rightarrow {\rm finite}}} G_{\alpha\beta}=0\cr
\lim_{\substack{\tiny M_p\rightarrow \infty\\ \Lambda_4\rightarrow {\rm finite}}} -\frac{G^{\alpha\beta}}{M_3^2}\pi_{\alpha\beta}= \frac{1}{2\Lambda_4^6}{\cal E}_4\ .
\end{eqnarray}
If instead gravity is {\it not} decoupled (i.e. $M_p$ is finite), one recovers the covariant quartic Galileaon of \cite{covgal} by noticing that the covariant Galileon is nothing else than $\Theta^{\mu\nu}\pi_{\mu\nu}$, as already mentioned. 

From now on we shall dub
\be
{\rm ``Slotheonic\ door"}:=G^{\alpha\beta}\pi_{\alpha\beta}\ ,
\ee
as this term will be the key interaction leading to Galilean equations of motion in certain gravity decoupling limits, similarly as in (\ref{similar}), as we shall see.
 
Now that we showed how to recover the quartic Galilean theory in flat spacetime from (\ref{curved}), we may ask whether the same would be possible for the quintic Galilean. In this case, in order to obtain four scalars into the scalar field equations of motion via integrating out gravity, one would need to use a stress tensor containing a term ``$(\pi_{\alpha\beta})^3$". This term is easily found in the action (\ref{ex}). Specifically, the stress-tensor associated to ${\cal L}_{\rm extra}$ is \cite{fabfour}
\be
M_5^5 T^{\epsilon\eta}=\frac{3}{2}{}^{**}R^{\eta\mu\epsilon\nu}\pi_\alpha\pi^\alpha\pi_{\mu\nu}+\frac{1}{2}g^{\epsilon\theta}\delta^{\eta\alpha\beta\gamma}{}_{\theta\mu\nu\sigma}\pi^\mu{}_\alpha\pi^\nu{}_\beta\pi^\sigma{}_\gamma\ .
\ee
The second term, substituted into the Slotheonic door, would indeed produce the same differential structure of the quintic Galileon suppressed by the scale $\Lambda_{\rm extra}\equiv (M_p^2 M_3^2 M_5^5)^{1/9}$.  However, in this case the scalar field equation would also contain the following extra term \cite{fabfour}
\be\label{pre}
M_5^5\frac{\delta{\cal L}_{\rm extra}}{\delta\pi}=3{}^{**}R^{\mu\nu\alpha\beta}\pi_{\mu\alpha}\pi_{\nu\beta}+\mbox{higher curvatures}\ .
\ee
In this case then, integrating out gravity in the scalar field equations would not only mean to solve for the Einstein tensor the Einstein equations, as in the Slotheonic door, but would also mean to solve for the full Riemann tensor. This can probably be done only in some specific cases and in any case the answer to the question of whether or not the quintic Galileon can be the obtained via (\ref{ex}) is left for future work. Nevertheless, it is interesting to note that, in a consistent DGP model, the term (\ref{ex}) would not appear and therefore, the quintic Galileon. This make us conjecture that the quintic galileon may indeed not be obtained from (\ref{curved},\ref{ex}). 

\subsection{The direct and derived origin of ${\cal E}_3$}\label{secdec}

Up to this point, we discovered that ${\cal E}_{4}$ is a {\it derived} equation emerging in a special decoupling limits of the theory ${\cal L}_{0,2}$, via the Slotheonic door ${\cal L}_2$. The equations ${\cal E}_{2,3}$ can be instead obtained by a direct variation of the minimally coupled parts of ${\cal L}_{2,3}$. 

We can now watch closer the tadpoles terms in ${\cal L}_1$. First of all, we notice that the Gauss-Bonnet term, in any consistent flat limit, should vanish. Therefore, the only term left that might still contribute to the Galilean theories in the flat spacetime limit, is the tadpole linear in curvatures, i.e. $\pi R$.

Let us focus again on the Slotheonic Lagrangian with in addition the tadpole term $M_1\pi R$, i.e. to the system
\be
{\cal L}_{\rm tot}=\frac{M_p^2}{2}\left(1+\frac{c_1}{M_p}\pi\right)R+\frac{1}{2 M_3^2}G^{\alpha\beta}\pi_{\alpha}\pi_\beta\ .
\ee
The scalar field equation is
\be\label{ricci}
M_1 R-\frac{G^{\alpha\beta}}{M_3^2}\pi_{\alpha\beta}=0\ .
\ee
The energy-momentum tensor associated to the linear curvature term in ${\cal L}_1$ reads \cite{pir}
\be\label{T}
T_{\alpha\beta}=-M_1 \pi G_{\alpha\beta}+M_1 (\nabla_{\alpha\beta}-g_{\alpha\beta}\square)\pi\ .
\ee
Therefore the total Einstein equations (neglecting terms of order ``$(\partial\pi)^2 R$" as before)
\be\label{solving}
\left(1+\frac{M_1}{M_p^2}\pi\right)G_{\alpha\beta}=\frac{M_1}{M_p^2}(\nabla_{\alpha\beta}-g_{\alpha\beta}\square)\pi+\frac{1}{M_3^2M_p^2}\left(-\pi_{\alpha\mu}\pi^{\alpha}_\nu+\pi_{\mu\nu}\pi_\alpha^{~\alpha}+\frac{1}{2}g_{\mu\nu}[\pi_{\alpha\beta}\pi^{\alpha\beta}-(\pi_\alpha^{~\alpha})^2]\right)\ .
\ee
One can check that a decoupling limit of this theory, reproducing the cubic Galileon via solving the Einstein equations (\ref{solving}) and substituting them into the Slotheonic door, is only possible for 
\be
M_1={\cal O} (M_p)\ .
\ee
In that case the decoupling will be exactly as in the quartic Galileon, i.e. for $M_p\rightarrow\infty$ and $\Lambda_4$ finite. However, in this case, the cubic Galileon would not be decoupled from the quadratic ($\square\pi$) and the quartic. This will be a very important property for the self-accelerating solutions of a consistent DGP model.

Finally then, if one is only interested in the covariantization of Galilean theory, up to the quartic theory, and it is also interested in decoupling the different Galilean theories, he/she may use the following covariant theory (we have re-labbeled the masses for simplicity)
\be\label{mincov}
{\cal A}_{\rm min I}=\int d^4x \sqrt{-g}\left[\frac{1}{2}(M_p^2+M_1 \pi)R+\frac{1}{2}\frac{G^{\alpha\beta}}{M_2^2}\partial_\alpha\pi\partial_\beta\pi+\frac{1}{M_3^3}(\partial\pi)^2\square\pi\right]\ .
\ee
Whereas the full extended version of a covariant Galilean action with redundancies in ${\cal E}_3$ and the higher curvature term (the Gauss-Bonnet tadpole), is
\be\label{covariant}
\!\!\!\!\!\!{\cal A}_{\rm cov}=\int d^4x \sqrt{-g}\left[\frac{1}{2}(M_p^2+\pi M_1)R-\frac{1}{2}\left(g^{\alpha\beta}-\frac{G^{\alpha\beta}}{M_2^2}\right)\partial_\alpha\pi\partial_\beta\pi+
\frac{1}{M_3^3}(\partial\pi)^2\square\pi+
\pi \frac{GB}{M_4}\right]\ .
\ee

\section{A consistent DGP model and the covariant Galileon}
 
In the DGP model, our Universe is a four dimensional brane embedded in a higher dimensional space (bulk). The peculiarity of this model is that the gravitational propagator is quasi-localized, i.e. it has a brane and bulk component. Explicitly, for a one extra dimension $r$, the original DGP action reads
\be\label{dgp}
A_{DGP}=\int d^5x\sqrt{-g}\left[R+r_c\delta(r){\cal R}+\ldots\right]\ ,
\ee
where $r_c$ is a length scale, $R\ ,\ {\cal R}$ are five and four-dimensional Ricci scalars and ``$\ldots$" are additional non-gravitational contributions.

Solutions with a self-accelerating four-dimensional cosmology has been found \cite{self}. This solution, may be reproduced locally by an effective ``non-galilean" theory formed by a general combination of (\ref{curved}) and the extra term (\ref{ex}), as proven in \cite{claudia}. However, these solutions, turned out to produce ghost instabilities at large scales \cite{bending,large}. Nevertheless, from an effective four dimensional perspective, the self acceleration may be locally (at distances far below the Hubble scale) reduced to a cubic Galileon sourcing gravity. At large scales however, the ghost-free Galilean solution would differ from the DGP ghost one. This might be due to the fact that the Lagrangian (\ref{dgp}) breaks explicitly covariance, as we shall discuss.

In \cite{spontaneous} we have shown that a consistent way to obtain a localized four-dimensional Ricci scalar is via the spontaneous localization mechanism. This mechanism, uses the simple fact that, whenever there is a brane, curvatures encode $\delta$-like terms that can be used to consistently (quasi)-localize \cite{quasi} (i.e. to consistently localize a four dimensional propagator on the brane), any field theory. In particular this can be used for gravity. In this context, it has been argued in \cite{spontaneous} that a consistent realization of a DGP-like scenario can be obtained from a Lovelock bulk lagrangian 
\be\label{non}
S_g=\frac{1}{2\ r_0}\int d^4x dr\sqrt{-g}\left[M_p^2 R+\alpha\ GB+\ldots\right],
\ee
where $r_0$ is the physical size of the extra dimension and $\alpha$ is a dimensionless coupling. Note that the realization (\ref{non}) of a consistent DGP-like model needs not to be unique, we will however only consider (\ref{non}) here.

Schematically, the linearized Gauss-Bonnet term would produce a Dirac delta function close to a brane such to localized a four-dimensional curvature at linearized level (the interested reader can see \cite{spontaneous}). It has then been shown in \cite{lov} that the dimensional reduction of $S_g$ would generically contain (\ref{covariant}), and in particular, it would generally contain the Slotheonic door. In this sense then a consistent DGP model would contain the Galilean theories up to ${\cal E}_4$. In other words, a generic dimensional reduction of a consistent DGP model (which is locally Galilean) should have the form (see for example \cite{coef})
\be\label{red1}
\!\!\!\!\!\!{\cal A}=\int d^4x \sqrt{-g}\left[\frac{1}{2}(M_p^2+\pi M_1)R-\frac{1}{2}\left(g^{\alpha\beta}-\frac{G^{\alpha\beta}}{M_2^2}\right)\partial_\alpha\pi\partial_\beta\pi+
\frac{1}{M_3^3}(\partial\pi)^2\square\pi+\pi \frac{GB}{M_4}\right]\ .
\ee
Although the knowledge of the precise mass scales $M_i$ is sensible to the compactification scheme, we can infer that the action (\ref{red1}) has the following structure
\be\label{red}
\!\!\!\!\!\!{\cal A}=\int d^4x \sqrt{-g}\left[\frac{1}{2}(M_p^2-c_1 M_p \pi)R-\frac{1}{2}\left(g^{\alpha\beta}-\frac{G^{\alpha\beta}}{M^2}\right)\partial_\alpha\pi\partial_\beta\pi+
\frac{c_4}{M^3}(\partial\pi)^2\square\pi+c_5\pi \frac{GB}{M}\right]\ ,
\ee
where $c_i={\cal O}(1)$ (that strictly depends on the compactification scheme) are dimensionless constants independent on $\alpha$ and $M\equiv M_2\propto \alpha^{-1}$. This result come from the fact that only the term $\pi R$ and the canonical kinetic term for $\pi$, comes from the five-dimensional Einstein-Hilbert Lagrangian, as shown in \cite{lov}. This is also in perfect compatibility with the fact that the only way to have a galilean theory out of the tadpole term $\pi R$, in the flat limit, is to have $M_1\sim M_p$, as discussed before. Finally, any other constant is coming from a dimensional reduction of the Gauss-Bonnet term \cite{lov}.

We saw however that in the action (\ref{red}) the cubic Galileon is redundant. We therefore expect that either $c_1$ or $c_4$ vanishes in some compactification scheme where the action (\ref{non}) would resemble the DGP model. Indeed, in \cite{coef} (see the first compactification scheme) it has been shown that, for co-dimension one, $c_4$ does vanish.
We therefore consider the restricted theory
 \be\label{redfinal}
\!\!\!\!\!\!{\cal A}_{\rm rest.}=\int d^4x \sqrt{-g}\left[\frac{1}{2}(M_p^2-c_1 M_p \pi)R-\frac{1}{2}\left(g^{\alpha\beta}-\frac{G^{\alpha\beta}}{M^2}\right)\partial_\alpha\pi\partial_\beta\pi+c_5\pi \frac{GB}{M}\right]\ .
\ee
Finally, note that, quantum mechanically, $M$ and $c_i$'s do not run up to the Planck scale \cite{non, sloth}.

\section{Self-acceleration}

As discussed before, the DGP action have a self-accelerating solution that propagates ghosts at very large scales. At a small scales, this solution may be interpreted as due to a light scalar field $\pi$ (the brane bending scalar \cite{bending}) sourcing gravity.

It has been showed that locally, i.e. for sub-horizon scales $|\vec x|^2H^2\ll1$ and around $t=0$, the self-accelerating solution of DGP can be described by a lagrangian of the form \cite{bending}
\be\label{jord}
L_{\rm DGP}\simeq \int d^4x \sqrt{-g}\left[\frac{1}{2}M^2_p\left(1-\frac{c_1}{M_p}\pi\right)R+{\cal L}_\pi\right]\ ,
\ee
where ${\cal L}_\pi$ is the Lagrangian of $\pi$ that is, at this level, unknown.

However, in the decoupling limit $M_p\rightarrow\infty$ and $\tilde\Lambda_1\equiv M_p c_1^{-1}$ finite, the equations of motion for $\pi$ are \cite{bending}
\be\label{scalar1}
\square\pi=\frac{a^3}{\tilde\Lambda_1^3}\left(\pi_{\alpha\beta}\pi^{\alpha\beta}-(\square\pi)^2\right)\ ,
\ee
where $a$ is a constant.

In the same limit instead, the Einstein equations are
\be\label{decoupled}
\left(1-\frac{\pi}{\tilde \Lambda_1}\right)G_{\alpha\beta}=-\frac{1}{\tilde\Lambda_1}\left(\pi_{\alpha\beta}-g_{\alpha\beta}\square\pi\right)\ .
\ee
Let us now suppose that an approximate DeSitter solution $G_{\alpha\beta}\simeq- 3H^2 \eta_{\alpha\beta}$ exists at sub-horizon scales, i.e. at scales in which $\vec {x}\cdot \vec{x}\ H^2\ll 1$ and around $t\sim 0$, where the index contraction has been performed by using the flat metric $\eta_{\alpha\beta}$ and $H$ is the constant Hubble scale. Moreover, we assume that masses of particles are constant in the Jordan frame of (\ref{jord}), as it should happen for the dimensional reduction of the DGP-like model (\ref{non}). In this way the Universe expansion in the Jordan frame will be physical.  Finally, in this region, all Christoffel symbols may be neglected while curvatures not. In this limit then the equations (\ref{scalar1},\ref{decoupled}) are invariant under Galilean transformations.

The solution we are looking for is therefore the next non trivial solution 
\be\label{pi}
\pi=\lambda^3 x_\mu x^\mu\ .
\ee
Plugging this ansatz into (\ref{scalar1}) we get
\be
\lambda^3=-\frac{\tilde\Lambda_1^3}{3 a^3}\ .
\ee
We can now solve (\ref{decoupled}) and find the following approximate sub-horizon solution
\be
H^2\simeq \frac{2}{3}\frac{\tilde\Lambda_1^2}{a^3}\ .
\ee
The question is now whether we will be able to find a similar solution whenever gravity is {\it not} decoupled.

We can follow the same steps as before, but having in mind that a consistent DGP model may only be found in the theory (\ref{non}). First of all, it is very interesting to note that $\tilde \Lambda_1\propto \Lambda_1$. Therefore, the decoupling limit discussed above reminds the decoupling limit of section \ref{secdec} that was necessary to obtain ${\cal E}_3$ by using the linear curvature tadpole in the Slotheonic door. However, here, there is a big difference: The use of the tadpole term needed to originate a cubic Galileon, necessarily lead to the appearance of an accompanying quartic Galileon. Therefore, the decoupling limit of a consistent DGP model (\ref{non}) will not, generically, be of the same form of (\ref{scalar1},\ref{decoupled}). 

We can now go beyond the decoupling limit and check whether an approximate DeSitter solution, at sub-horizon scales, still exists. In this respect we will consider the theory  (\ref{redfinal}). Again at this scales, curvatures will not vanish but the contribution of the Christoffel symbols can be neglected. In this limit the $\pi$ equation of motion of (\ref{redfinal}), is locally Galilean invariant, as already noticed in \cite{sloth}\footnote{Note that this would not be the case in the presence of the term (\ref{ex}), as explained in \cite{sloth}.}. We then have good reasons to expect that the solution (\ref{pi}), is still a solution of the non-decoupled system.

Let us now suppose that, as before, an approximate DeSitter solution $H\simeq\rm const$, exists. Then, at sub-horizon scales any term of the form ``$\partial\pi\partial\pi R$", where $R$ is a generic curvature, can be neglected with respect to ``$(\square\pi)^2$". We will then assume for the time being that $\frac{c_1\pi}{M_p}\ll 1$ \footnote{This must be, as this tadpole term is coming from the expansion of a dilaton type interaction $\sim e^{\frac{\pi}{M_p}}R$.}, the canonical stress tensor and the Gauss-Bonnet contributions are negligible.  We will check all these assumptions a posterior. 

Summarizing we will now consider the following approximate Lagrangian
\be\label{dgp-like2}
\!\!\!\!\!\!{\cal A}_{\rm rest.}\simeq\int d^4x \sqrt{-g}\left[\frac{1}{2}(M_p^2-c_1 M_p \pi)R-\frac{1}{2}\left(g^{\alpha\beta}-\frac{G^{\alpha\beta}}{M^2}\right)\partial_\alpha\pi\partial_\beta\pi\right]\ .
\ee

Under these conditions, the gravity equations reads\footnote{We remind the reader that the canonical stress tensor has been neglected. This assumption will be checked at posterior.}
\be\label{G}
G_{\alpha\beta}\simeq-\frac{c_1}{M_p}\left(\pi_{\alpha\beta}-g_{\alpha\beta}\square\pi\right)+\frac{1}{M_p^2M^2}\left[-\pi_{\alpha\mu}\pi^{\alpha}_\nu+\pi_{\mu\nu}\pi_\alpha^{~\alpha}+\frac{1}{2}g_{\mu\nu}\left(\pi_{\alpha\beta}\pi^{\alpha\beta}-(\pi_\alpha^{~\alpha})^2\right)\right]\ ,
\ee
and the scalar equations
\be
\left(g^{\alpha\beta}-\frac{G^{\alpha\beta}}{M^2}\right)\pi_{\alpha\beta}\simeq-c_1 M_p R\ .
\ee
Plugging the ansatz discussed before for the metric and the scalar, we find from the scalar field equation
\be\label{soll}
\lambda^3\simeq-\frac{3}{2}\frac{c_1 M_p H^2}{1+3\frac{H^2}{M^2}}\ .
\ee
We can plug this result into the gravity equations (\ref{G}). What is interesting to note is that the quartic Galilean $(t,t)$ component vanishes. Therefore, for this component, the system is exactly like (\ref{decoupled}). In this case then, the $(t,t)$ component (the Hubble equation) results in
\be\label{hsol}
H^2\simeq -2\frac{c_1}{M_p}\lambda^3\Rightarrow H^2\simeq M^2\frac{3c_1^2-1}{3},
\ee
where the last equality have been obtained by using the solution (\ref{soll}). The solution (\ref{hsol}) is only valid for the range
\be\label{range}
c_1^2\geq\frac{1}{3}\ .
\ee
We now need to check whether our previous assumptions are correct. It is easy to check that all the terms we neglected a priori are indeed negligible for the solution (\ref{hsol}) with the range (\ref{range}) and for $M_p\gg M$. Note also the interesting property that a very tiny cosmological constant can be obtained even for a large mass $M$ by going closer and closer to the critical point $c_1^2=\frac{1}{3}$.

Whether the range of parameter (\ref{range}) corresponds to a physical compactification of the DGP-like model (\ref{non}) is left for future investigation. However, at this level, we see no impediments for this to happen.
Even more, without working out the details, if, as we expect, the self-accelerating solution exists, it can be physically understood by noticing that the approximate solution (\ref{pi}), would imply a slight bending of the brane at horizon scales. This same bending, would then generate a local acceleration via a ``brane-stretching" effect, just as in the case of non-homogeneous mirage cosmologies \cite{mirage}.

Let us now discuss upon a possible problem arising from the tadpole term $\pi R$ in the effective theory (\ref{covariant}). It would indeed seem that the effective Newtonian constant ($G_N\sim M_p^{-2}(1-\frac{\pi}{\tilde\Lambda_1})$) could become negative whenever $\pi>\tilde\Lambda_1$. From the effective field theory point of view, if we were coming from the positive Newtonian side ($\pi<\tilde\Lambda_1$) we would never evolve into the negative side as, in any spacetime point in which $\pi\sim \tilde\Lambda_1$, a Black Hole would form much before \cite{gia}. The question is whether there could be any solution starting with $\pi>\tilde\Lambda_1$, i.e. whether there could be a background with graviton-ghost propagation. If the effective theory is a dimensionally reduced theory of a consistent theory of gravity, as in (\ref{non}), we would expect this not to happen. In other words, the effective field theory (\ref{dgp-like2}) can only be trusted for ``small'' $\pi$, for large values the full five dimensional theory must be used.

Finally, it interesting to note that, since we are on a quasi-DeSitter universe and in the high friction regime, the perturbative strong coupling of this theory is $\sim M_p$, for {\it any} choice of $M$ \cite{bending,non,high}.

 \section{Conclusions}
 
 In this paper we showed that the Galilean theories up to the quartic in flat spacetime are obtained in the (gravity) decoupling limit of general Slotheonic theories
 \be\label{final}
\!\!\!\!\!\!{\cal A}_{\rm cov}=\int d^4x \sqrt{-g}\left[\frac{1}{2}(M_p^2+\pi M_1)R-\frac{1}{2}\left(g^{\alpha\beta}-\frac{G^{\alpha\beta}}{M_2^2}\right)\partial_\alpha\pi\partial_\beta\pi+
\frac{1}{M_3^3}(\partial\pi)^2\square\pi+
\pi \frac{GB}{M_4}\right]\ .
\ee

The Galilean terms in flat space are simply obtained when the Einstein equations are substituted into the Slotheonic door 
\be
{\rm ``door"}\equiv G^{\alpha\beta}\nabla_{\alpha\beta}\pi\ ,
\ee
in the decoupling limits
\be\label{special}
G_{\alpha\beta}\rightarrow 0\ ,\cr
\frac{G_{\alpha\beta}}{M_2^2}\rightarrow {\rm finite}\ .
\ee
In this theory, the Galilean equation of motion ${\cal E}_{4}$ is {\it derived}.

The Slotheonic door can be super-symmetrized consistently in the framework of new-minimal supergravity \cite{sugra}. Therefore, a consistent supersymmetrization of all Galilean theories up to the quartic order may only be done in the context of supergravity and in the special decoupling limit (\ref{special}). 

In this framework, we argued that the effective four-dimensional field theory of a consistent DGP-like model \cite{spontaneous} should be of the form (\ref{final}) with specific coefficients related to the chosen compactification scheme. 
Within this context, we showed that solutions with local (sub-horizon) DeSitter acceleration exist and that they could reproduce the observed late time Universe acceleration. From the DGP point of view, the scalar source that accelerate the Universe is interpreted as the brane bending scalar mode. 
One may however ask what the cosmological solution would be at super-horizon scales. Where we could certainly assert that the solution will not be homogeneous and isotropic, we could not use the effective Galilean theory (\ref{final}) to describe it. Indeed, the theory becomes more and more ``five dimensional" at larger and larger scales. This is a peculiarity of the quasi-localization mechanism \cite{quasi} used in this paper in order to obtain (\ref{final}). We leave the study of the super-horizon solution for future work.

Finally, whether or not the self-accelerated solution found in this paper is stable against scalar, tensor and vector perturbations is left for future work. Nevertheless, as the five-dimensional Einstein-Gauss-Bonnet theory does not propagate ghost degrees of freedom, we expect here the absence of the scalar ghost instabilities that plagued the self-accelerating solutions of the original DGP model \cite{self}.

 \section*{Acknowledgements}
 I wish to thank Gia Dvali for discussion on the decoupling limit of gravity. I wish to thank Alex Kehagias and Yuki Watanabe for carefully reading the first draft of this paper. I would like to thank Claudia De Rham for discussions on the Gauss-Bonnet term. Finally, I would like to thank Luca Martucci for useful suggestions and Yuki Watanabe and Nico Wintergerst for important discussions. I am supported by Alexander Von Humboldt Foundation.

 \end{document}